\documentclass[12pt]{iopart}
\usepackage{setstack}
\newcommand{\eo}{\epsilon(\omega)}
\newcommand{\mo}{\mu(\omega)}
\newcommand{\F}{\mathcal{F}}
\newcommand{\kb}{k_{\mathrm{B}}}
\newcommand{\Li}[2]{\mathrm{Li}_{#1}\left(#2\right)}
\newcommand{\Ei}[1]{\mathrm{Ei}\left(#1\right)}
\newcommand{\Dt}{\Delta_{T}}
\begin{document}
\title{Casimir entropy for magnetodielectrics}
\author{G.~L.~Klimchitskaya$^{1,2}$ and C.~C.~Korikov$^2$}
\address{$^1$~Central Astronomical Observatory at Pulkovo of the Russian Academy of Sciences, St. Petersburg, 196140, Russia}
\address{$^2$~Institute of Physics, Nanotechnology and Telecommunications, St. Petersburg State Polytechnical University, St. Petersburg, 195251, Russia}
\ead{\mailto{g.klimchitskaya@gmail.com}, \mailto{korikov.constantine@spbstu.ru}}

\begin{abstract}
    We find the analytic expressions for the Casimir free energy, entropy and
    pressure at low temperature in the configuration of two parallel plates
    made of magnetodielectic material. The cases of constant and frequency-dependent
    dielectic permittivity and magnetic permeability of the plates are considered.
    Special attention  is paid to the account of dc conductivity. It is shown that
    in the case of finite static dielectric permittivity and magnetic permeability
    the Nernst heat theorem for the Casimir entropy is satisfied. If the dc
    conductivity is taken into account, the Casimir entropy goes to a positive
    nonzero  limit depending on the parameters of a system when the temperature vanishes, i.e., the Nernst theorem
    is violated. The experimental situation is also discussed.
\end{abstract}

\maketitle

\section{Introduction}\label{sec:intro}
    The Casimir force\cite{r01} acts between closely spaced material bodies. It is caused by
    the zero-point and thermal fluctuations of the electromagnetic field.
    The Casimir force is of the same physical nature as the van der Waals force,
    but it acts at larger separation distances (typically  starting from a few
    nanometers to a few micrometers), where the effects of relativistic
    retardation become essential. The Casimir effect  finds  numerous and diverse
    applications ranging from  condensed matter physics and nanotechnology to
    quantum field theory, gravitation and cosmology\cite{r02,r03,r04}. Specifically, in nanotechnology
    the Casimir force can lead to a stiction and a loss of functionality in
    microdevices\cite{r05}, on the one hand, and plays a useful role by actuating microdevices\cite{r06},
    on the other hand.

    Advances in microfabrication techniques allowed to perform precise
    measurements of the Casimir force and its gradient at separations up to a few
    hundred nanometers at room and lower temperature (see reviews\cite{r07,r08,r09} and some of
    later experiments\cite{r10,r11,r12,r13,r14,r15}). The measurement results
    have been compared with theoretical predictions of the Lifshitz theory\cite{r16} of the van der Waals and Casimir forces.
    This comparison resulted in a big surprise. The Lifshitz theory uses  the
    frequency-dependent dielectric permittivity  $\eo$ of interacting
    bodies as an input data for calculation. It turned out that  if the low-frequency
    behavior of $\eo$ is described by the most precise Drude model,
    taking into account the relaxation properties of conduction electrons in metals,
    the theoretical results are excluded by the data at a high confidence level\cite{r04,r07,r12}.
    If, alternatively, the low-frequency behavior of $\eo$ is described by the
    nondissipative plasma model, the theoretical results are consistent with the
    measurement data. (There is only one experiment\cite{r17}, which is claimed to be in
    agreement with the Drude model; it  measures, however, not the Casimir force,
    but up to an order of magnitude larger force of unclear origin and subtracts the major part
    of it by means of a fitting procedure which was critically discussed in the
    literature\cite{r18,r19}.)

    In parallel with the experimental progress, the analytic behavior of the Casimir
    free energy and entropy at low temperature has been found in the framework of the Lifshitz
    theory for the case of  two metallic test bodies\cite{r20,r21,r22}. Here also a big surprise
    has been found supposedly connected with the above puzzle in theory-experiment comparison.
On the one hand,
    for metal plates with perfect  crystal lattices, the Casimir entropy calculated by using
    the Lifshitz theory combined  with the Drude model was shown to go  to a nonzero
    negative limit depending on the parameters of the system when the temperature
    vanishes\cite{r20,r21,r22}. This means the violation of the third
    law of thermodynamics (the Nernst heat theorem). If the dielectric permittivity
    of the plasma model is used, the Casimir entropy was shown to go to the zero limit
    when temperature goes to zero, i.e., the Nernst heat theorem is satisfied.
On the other hand, at large separations, where the Casimir force
is classical, the plasma model was shown \cite{22a} to violate
the Bohr-van Leeuwen theorem, whereas the Drude model
is consistent with it.
     Later
    on, the thermodynamic inconsistency was also revealed for dielectric plates
    with taken into account conductivity at a constant current (the dc conductivity).
    It was shown\cite{r23,r24,r25} that in this case the Casimir entropy
    goes to a nonzero positive limit depending on the parameters of a system
    when the temperature vanishes. The experimental data of several
    experiments using dielectric materials were found in agreement with theoretical
    predictions of the Lifshitz theory only if the contribution of free charge
    carriers into the dielectric permittivity is omitted\cite{r04,r07,r09,r11,r26,r27,r28}.

    Many attempts were undertaken in order to solve the problem of thermodynamic
    inconsistency of the Lifshitz theory. Specifically, it was argued\cite{r29} that
    for metallic crystal lattices with some concentration of impurities there is some
    small but nonzero residual relaxation at zero temperature. As a result, at very
    low temperature the Casimir entropy undergoes an abrupt increase to zero
    and the Nernst heat theorem is formally satisfied. The problem remains, however,
    unsolved for metals with perfect crystal lattices because the perfect lattice is a truly
    equilibrium system with the nondegenerate dynamical state of lowest energy.
    Thus, the Nernst heat theorem should be satisfied in this case. Another attempt to avoid
    thermodynamic inconsistency for dielectrics\cite{r30} and metals\cite{r31} goes
    beyond the Lifshits theory and modifies it by taking into account screening
    effects and diffusion currents of free charge carriers. It was shown, however,
    that for metals with perfect crystal lattices, as well as for those dielectrics
    whose concentration of free charge carriers does not go to zero with vanishing
    temperature (for such dielectric materials conductivity goes to zero due to vanishing mobility),
    the Nernst heat theory remains to be violated in the modified theory\cite{r32,r33,r34}.
    Furthermore, the modified theory turned out to be excluded by the measurement
    data of experiments with metallic test bodies at the same high level of confidence
    as the standard Drude model\cite{r33, r34}. For dielectric test bodies the modified
    theory was found to be consistent with the data of experiment\cite{r27},
    but still excluded by the experiment\cite{r26}.

    In this paper, we investigate the low-temperature behavior of the Casimir
    free energy, entropy and pressure for two parallel plates made of a magnetodielectric
    material. The Lifshitz theory was generalized for the case of magnetic materials
    long ago\cite{r35}, and this subject recently received considerable
    attention in the literature\cite{r36,r37,r38,r39}. It became especially
    topical after successful measurements of the gradient of the Casimir force
    between magnetic materials\cite{r40,r41,r42}, which again excluded the Drude
    model and turned out to be consistent with the plasma model. Here, we show
    that for magnetodielectric plates the low-temperature behavior of all physical
    quantities is typically the same as for dielectric ones, and only the respective
    coefficients are generalized due to the presence of an additional
    parameter --- the magnetic permeability $\mo$. As a result, we have proven
    that for magnetodielectric plates with omitted dc conductivity the  Nernst
    heat theorem is satisfied and, if the dc conductivity is included, the  Nernst
    theorem is violated. This result is obtained for both constant
    and frequency-dependent magnetic permeability.

    The paper is organized as follows. Section~\ref{sec:formalism} contains brief
    presentation of the formalism adapted for use in perturbation expansions.
    In section~\ref{sec:const} we derive the asymptotic expressions for
    the Casimir free energy, entropy and pressure at low temperature
    for magnetodielectrics with constant dielectric permittivity and magnetic
    permeability. Section~\ref{sec:freq} is devoted to the account of frequency
    dependence. In section~\ref{sec:concl} we formulate our conclusions.

\section{Basic formalism}\label{sec:formalism}
    We consider two parallel thick magnetodielectric plates (semispaces)
    characterized by the frequency-dependent dielectric permittivity $\epsilon$
    and magnetic permeability $\mu$ at temperature $T$ in thermal equilibrium
    with an environment. In terms of dimensionless variables, the Lifshitz
    formula for the free energy of the Casimir interaction per unit area
    of the plates is given by\cite{r04,r16}
    \begin{equation}\label{eq:01}
        \F(a,T) = \frac{\hbar c \tau}{32 \pi^2 a^3} \sum_{l=0}^{\infty}\phantom{}^{'}
        \int_{\tau l}^{\infty} y\,\rmd y
        \left[
            \ln\left(1-r^2_{\mathrm{TM}}\rme^{-y}\right)+\ln\left(1-r^2_{\mathrm{TE}}\rme^{-y}\right)
        \right].
    \end{equation}
    Here, $a$ is the separation distance between the plates and the dimensionless
    quantity $\tau = 4 \pi \kb a T/ (\hbar c)$ where $\kb$ is the Boltzmann constant.
    The prime near the summation sign means that  the term with $l=0$ is divided by two.
    The reflection coefficients for the two independent polarizations of the electromagnetic
    field, transverse magnetic (TM) and transverse electric (TE), are given by
    \begin{equation}\label{eq:02}
    \eqalign{
        r_{\mathrm{TM}} &=  r_{\mathrm{TM}} (\rmi\zeta_l,y) = \frac{\epsilon_l y-\sqrt{y^2+\zeta_l^2(\epsilon_l\mu_l-1)}}{\epsilon_l y+\sqrt{y^2+\zeta_l^2(\epsilon_l\mu_l-1)}},\\
        r_{\mathrm{TE}} &=  r_{\mathrm{TE}} (\rmi\zeta_l,y) = \frac{\mu_l y-\sqrt{y^2+\zeta_l^2(\epsilon_l\mu_l-1)}}{\mu_l y+\sqrt{y^2+\zeta_l^2(\epsilon_l\mu_l-1)}}.
    }
    \end{equation}
    In these expressions, we use the notations $\epsilon_{l}\equiv \epsilon(\rmi \xi_{l})$, $\mu_{l}\equiv \mu(\rmi \xi_{l})$, $\xi_{l} = 2\pi \kb Tl /\hbar$ and $\zeta_l=\xi_{l}/\xi_{c}=l \tau$, where $\xi_{c} \equiv c/(2a)$, are the dimensional and dimensionless Matsubara frequencies, respectively. Note also that the dimensionless variable $y=2a(k_{\perp}^2+\xi_l^2/c^2)^{1/2}$ where $\mathbf{k_{\perp}}$ is the projection of the wave vector on the plane of plates.

    By applying the Abel-Plana formula\cite{r04}
    \begin{equation}\label{eq:03}
        \sum_{l=0}^{\infty}\phantom{}^{'} F(l) = \int_{0}^{\infty} F(t)\,\rmd t + \rmi \! \int_{0}^{\infty}\rmd t \,\frac{F(\rmi t)-F(-\rmi t)}{\rme^{2\pi t}-1},
    \end{equation}
    one can rearrange~\eref{eq:01} to the for form\cite{r04}
    \begin{equation}\label{eq:04}
         \F(a,T) = E(a)+\Dt \F(a,T),
    \end{equation}
    where the first term on the right-hand side is the Casimir energy per unit area at zero temperature
    \begin{equation}\label{eq:05}
    \eqalign{
         E(a) = \frac{\hbar c}{32 \pi^2 a^3} \int_{0}^{\infty} \rmd\zeta \, F(\zeta),\\
         F(\zeta)\equiv \int_{\zeta}^{\infty} \rmd y \, f(\zeta,y),\\
         f(\zeta,y) \equiv y \left\{ \ln \left[ 1- r^2_{\mathrm{TM}} (\rmi \zeta,y) \rme^{-y} \right] + \ln \left[ 1- r^2_{\mathrm{TE}} (\rmi \zeta,y) \rme^{-y} \right] \right\}.
    }
    \end{equation}
    The second term on the right-hand side of~\eref{eq:04} has the meaning of the thermal correction
    \begin{equation}\label{eq:06}
        \Dt \F(a,T)= \frac{\rmi \hbar c \tau}{32 \pi^2 a^3} \int_{0}^{\infty} \rmd t \, \frac{F(\rmi t \tau)-F(-\rmi t \tau)}{\rme^{2\pi t}-1}.
    \end{equation}

    In a similar way, the Casimir pressure can be represented in the form
    \begin{equation}\label{eq:07}
        P(a,T) = P_{0}(a)+\Dt P(a,T),
    \end{equation}
    where the Casimir pressure at $T=0$ is \cite{r04}
    \begin{equation}\label{eq:08}
    \eqalign{
         P_{0}(a)&=-\frac{\hbar c}{32 \pi^2 a^4} \int_{0}^{\infty} \,\rmd \zeta \Phi(\zeta),\\
         \Phi(\zeta) &\equiv \int_{\zeta}^{\infty} \rmd y \, g(\zeta,y),\\
         g(\zeta,y) &\equiv y^2
        \left[
            \frac{r^2_{\mathrm{TM}}(\rmi \zeta,y)}{\rme^{y}-r^2_{\mathrm{TM}}(\rmi\zeta,y)}
            +
            \frac{r^2_{\mathrm{TE}}(\rmi \zeta,y)}{\rme^{y}-r^2_{\mathrm{TE}}(\rmi\zeta,y)}
        \right],
    }
    \end{equation}
    and the thermal correction to it is given by\cite{r04}
    \begin{equation}\label{eq:09}
        \Dt P(a,T)= -\frac{\rmi \hbar c \tau}{32 \pi^2 a^4} \int_{0}^{\infty} \rmd t \, \frac{\Phi(\rmi t \tau)-\Phi(-\rmi t \tau)}{\rme^{2\pi t}-1}.
    \end{equation}
\section{Casimir free energy, entropy and pressure at low temperature for constant $\epsilon$ and $\mu$}\label{sec:const}

    In this section we describe the magnetodielectric plates by their static dielectric
    permittivity $\epsilon_0 \equiv \epsilon(0)$ and magnetic permeability $\mu_0 \equiv \mu(0)$.
    In fact at sufficiently low $T$ the Matsubara frequencies, giving the leading
    contribution to the thermal corrections~\eref{eq:06}
    and~\eref{eq:09} for insulating material with no free
    charge carriers, belong to the region where $\epsilon$ is almost constant and equal
    to $\epsilon_0$. Because of this, the qualitative behavior of the Casimir free
    energy at low temperature for dielectric materials described by constant
    and frequency-dependent $\epsilon$ is similar (only a power in the
    power-type law can be different\cite{r23,r25}).

    The account of dc conductivity in the dielectric permittivity and
    frequency-dependence of the magnetic permeability are discussed in section~\ref{sec:freq}.

\subsection{Thermal correction to the Casimir energy in the lowest order}
    Now we assume that $\zeta \ll 1$ and expand the function $f(\zeta,y)$ defined in~\eref{eq:05} in powers of this small parameter
    \begin{equation}\label{eq:10}
        \hspace*{-1cm}
        f(\zeta,y)=y \sum_{k=1,2} \left[ \ln\left(1-r^2_{0,k}\rme^{-y}\right) + \frac{2 \alpha_k \beta}{y \left(1+\alpha_k\right)^2} \frac{r_{0,k}\rme^{-y}}{1-r_{0,k}^2\rme^{-y}} \zeta^2 + O(\zeta^4)\right].
    \end{equation}
    Here, the following notations are introduced:
    \begin{equation}\label{eq:11}
        \alpha_1 = \epsilon_0, \qquad \alpha_2 = \mu_0, \qquad  \beta=\epsilon_0\mu_0-1, \qquad r_{0,k} = \frac{\alpha_k-1}{\alpha_k+1}.
    \end{equation}
    Integrating equation~\eref{eq:10} in accordance to~\eref{eq:05}, one obtains
    \begin{equation}\label{eq:12}
        \hspace*{-1cm}
        F(\zeta) = - \sum_{k=1,2} \left[ \Li{3}{r_{0,k}^2} - \frac{\zeta^2}{2} \ln\left(1-r_{0,k}^2\right) - \zeta^2 \sum_{n=1}^{\infty} r_{0,k}^{2n}\Ei{-n\zeta}\right]+O(\zeta^3),
    \end{equation}
    where $\Li{n}{z}$ is the polylogarithm function and $\Ei{z}$ is the exponential integral function.

    Now we take into account the identity\cite{r43}
    \begin{equation}\label{eq:13}
        \Ei{\rmi x} - \Ei{-\rmi x} = - \rmi \pi + O(x)
    \end{equation}
    and calculate the quantity $F(\rmi t \tau)-F(-\rmi t \tau)$ using equation~\eref{eq:12}, where we put $\zeta = t \tau \ll1$:
    \begin{equation}\label{eq:14}
        F(\rmi \tau t) - F(-\rmi \tau t) = \rmi \pi \frac{\beta}{2} \sum_{k=1,2} r_{0,k}t^2\tau^2 + O(\tau^3).
    \end{equation}
    Substituting this expression in~\eref{eq:06} and performing the integration with respect to $t$, one finds
    \begin{equation}\label{eq:15}
        \Dt \F(a,T) = - \frac{\hbar c}{32 \pi^2 a^3} \left[ \frac{\zeta(3) \left(\epsilon_0\mu_0-1\right)^2}{4 \pi^2 \left(\epsilon_0+1\right)\left(\mu_0+1\right)} \tau^3 + O(\tau^4) \right],
    \end{equation}
    where $\zeta(z)$ is the Riemann zeta function.
    This is the lowest order thermal correction to the Casimir energy for magnetodielectric plates with constant $\epsilon$ and $\mu$. For $\mu_0=1$ it coincides with the previously obtained result\cite{r24}.

    From~\eref{eq:15} it is straightforward to get the analytic behavior of the Casimir entropy per unit area at low temperature
    \begin{equation}\label{eq:16}
        S(a,T) = - \frac{\partial \Dt \F(a,T)}{\partial T} = \frac{3\zeta(3) \kb^3}{2 \pi (\hbar c)^2} \frac{(\epsilon_0\mu_0-1)^2}{(\epsilon_0+1)(\mu_0+1)} T^2 + O(T^3).
    \end{equation}
    Here, we returned to the dimensional temperature $T$.
    As can be seen from this expression, the Casimir entropy goes to zero when the temperature vanishes, i.e., the Nernst heat theorem is satisfied for the magnetodielectric plates.

    Note that the right-hand side of~\eref{eq:15} does not depend on $a$ because $\tau \sim a$. This does not allow to obtain from~\eref{eq:15} the asymptotic behavior of the thermal correction to the Casimir pressure at low temperature. It is also difficult to find the next, fourth order, term of the perturbation expression in~\eref{eq:15} because all powers of the expansion of $f(\zeta,y)$ in equation~\eref{eq:10} contribute to its coefficient. In the next subsection the perturbation asymptotic expansion for the Casimir pressure is obtained in a direct way starting from equation~\eref{eq:09}. This makes possible to find the next terms in the asymptotic expressions~\eref{eq:15} and~\eref{eq:16} for the thermal correction to the Casimir energy and for the Casimir entropy, respectively.
\subsection{Thermal correction to the Casimir pressure in the lowest order}
    We begin with the function $g(\zeta,y)$ defined in~\eref{eq:08} and expand it in powers of a small parameter $\zeta$
    \begin{equation}\label{eq:17}
        g(\zeta,y) = \sum_{k=1,2} \left[ y^2 \frac{r^2_{0,k}}{\rme^y-r_{0,k}^2}-\frac{2\alpha_k\beta}{\left(1+\alpha_k\right)^2}\frac{r_{0,k}\rme^{-y}}{\left(1-r_{0,k}^2\rme^{-y}\right)^2}\zeta^2\right]+ O(\zeta^4).
    \end{equation}
    Then we integrate~\eref{eq:17} in accordance with~\eref{eq:08} and obtain
    \begin{equation}\label{eq:18}
        \hspace*{-1cm}
        \Phi(\zeta) = \sum_{k=1,2} \left[ 2\Li{3}{r^2_{0,k}} - \frac{\beta}{2} r_{0,k} \zeta^2+\frac{\alpha_k\left(\beta-1\right)\sqrt{\beta+1}+2}{6}\zeta^3\right] + O(\zeta^4).
    \end{equation}
    Using this expression, the difference of functions $\Phi$ entering equation~\eref{eq:09} takes the form
    \begin{equation}\label{eq:19}
        \Phi(\rmi t \tau) -\Phi(-\rmi t \tau) = - \rmi t^3 \tau^3 \frac{(\epsilon_0+\mu_0)(\epsilon_0\mu_0-2)\sqrt{\epsilon_0\mu_0}+2}{3}+O(\tau^4).
    \end{equation}
    Substituting this in~\eref{eq:09} and integrating with respect to $t$, one finds
    \begin{equation}\label{eq:20}
        \Dt P(a,T) = - \frac{\hbar c}{32 \pi^2 a^4} \left[\frac{(\epsilon_0+\mu_0)(\epsilon_0\mu_0-2)\sqrt{\epsilon_0\mu_0}+2}{720}\tau^4+O(\tau^5)\right].
    \end{equation}
    This result presents the low-temperature behavior of the thermal correction
    to the Casimir pressure for magnetodielectric plates. In the
    absence of magnetic properties ($\mu_0=1$) it coincides with the known result\cite{r24}.

    Equation~\eref{eq:20} allows to find the more exact than~\eref{eq:15} and~\eref{eq:16} asymptotic expansions for the thermal correction to the Casimir energy and Casimir entropy. For this purpose we use the relationship
    \begin{equation}\label{eq:21}
        \Dt P(a,T) = - \frac{\partial \Dt \F(a,T)}{\partial a}
    \end{equation}
    and by integrating with respect to $a$ find
    \begin{equation}\label{eq:22}
        \hspace*{-1cm}
        \Dt \F(a,T) = \frac{\hbar c}{32 \pi^2 a^3} \frac{(\epsilon_0+\mu_0)(\epsilon_0\mu_0-2)\sqrt{\epsilon_0\mu_0}+2}{720}\tau^4 + \mathrm{const}+O(\tau^6).
    \end{equation}

    An independent on $a$ constant was already determined in~\eref{eq:15}. Thus, combining~\eref{eq:15} and~\eref{eq:22}, we obtain a more exact  asymptotic expansion for the thermal correction to the Casimir energy of magnetodielectric plates
    \begin{equation}\label{eq:23}
    \eqalign{
        \Dt \F(a,T) = &-\frac{\hbar c}{32 \pi^2 a^3} \left[ \frac{\zeta(3)\left(\epsilon_0\mu_0-1\right)^2}{4\pi^2\left(\epsilon_0+1\right)(\mu_0+1)} \tau^3 \right.\\ &\left.-\frac{(\epsilon_0+\mu_0)(\epsilon_0\mu_0-2)\sqrt{\epsilon_0\mu_0}+2}{720}\tau^4+O(\tau^5)\right].
    }
    \end{equation}
    The respective more exact expression for the Casimir entropy per unit area at low temperature is given by
    \begin{equation}\label{eq:24}
    \eqalign{
        S(a,T)= &\frac{\kb^3 T^2}{(\hbar c)^2} \left\{ \frac{3\zeta(3)}{2\pi} \frac{(\epsilon_0\mu_0-1)^2}{(\epsilon_0+1)(\mu_0+1)} \right. \\
        &\left.- \frac{2 \pi^2 \kb a}{45 \hbar c} \left[ (\epsilon_0+\mu_0)(\epsilon_0\mu_0-2)\sqrt{\epsilon_0\mu_0}+2\right]T + O(T^2)\right\}.
    }
    \end{equation}
    The second term in this asymptotic expression depends both on the temperature and on separation distance.
\section{Account for frequency dependence}\label{sec:freq}
    As was mentioned in the beginning of section~\ref{sec:const}, the frequency dependence of the dielectric
     permittivity of dielectric materials plays only a minor role in the behavior of the Casimir free energy
     and entropy at low temperature\cite{r23,r25}. In this section we investigate the impact of the frequency
     dependence of the magnetic permeability. Then we consider what happens  when the magnetodielectric material
     contains some small fraction of free charge carriers which is the case for any real dielectric at nonzero
     temperature.
\subsection{Frequency-dependent magnetic permeability}
    The dependence of the magnetic permeability on the frequency along the imaginary frequency axis is
    described by the Debye formula\cite{r44}
    \begin{equation}\label{eq:25}
        \mu(\rmi \xi) = 1 + \frac{\mu_0-1}{1+\frac{\xi}{\omega_m}},
    \end{equation}
    where $\omega_m$ is some characteristic frequency which is different for different magnetic materials.
    Using the dimensionless variables introduced in section~\ref{sec:formalism}, one has
    \begin{equation}\label{eq:26}
        \mu(\rmi \xi) = 1 + \frac{\mu_0-1}{1+\kappa_m \zeta},
    \end{equation}
    where $\kappa_m = c/(2a \omega_m)$.

    Now we expand in powers of $\zeta$ the function $f(\zeta,y)$ defined in~\eref{eq:05},
    where $\epsilon(\rmi \xi) = \epsilon_0$ remains constant but $\mu(\rmi \xi)$ is given by~\eref{eq:26}.
    As a result, we again arrive to the asymptotic expansion~\eref{eq:10} with the following additional
    contribution to the right-hand side:
    \begin{equation}\label{eq:27}
        \widetilde{f}(\zeta,y) = 4 \frac{\kappa_m r^2_{0,2}}{\mu_0+1} \frac{y}{\rme^y-r^2_{0,2}} \zeta + \frac{4 \kappa_m^2 r_{0,2}^2}{\left(\mu_0+1\right)^2} \frac{\left(r_{0,2}^2-3 \rme^y\right)y}{\left(\rme^y-r^2_{0,2}\right)^2}\zeta^2+O(\zeta^3).
    \end{equation}
    It is seen that now the asymptotic expansion of the function $f(\zeta,y)$ contains the first order
    term in $\zeta$. In the case of constant magnetic permeability $\kappa_m=0$ and $\widetilde{f}(\zeta,y)=0$.

    We substitute the sum $f(\zeta,y)+\widetilde{f}(\zeta,y)$ to the definition of the function
    $F(\zeta)$ in~\eref{eq:05} in place of $f(\zeta,y)$ and in the first perturbation order arrive at
    \begin{equation}\label{eq:28}
        F(\zeta) = - \sum_{k=1,2} \Li{3}{r^2_{0,k}} + \frac{4 \kappa_m}{\mu_0+1} \Li{2}{r_{0,2}^2} \zeta + O(\zeta^2).
    \end{equation}

    Note that the term of the order of $\zeta^2$ in~\eref{eq:28} does not contribute to the
    difference $F(\rmi t \tau)-F(-i t \tau)$. Then with the help of~\eref{eq:28} we obtain
    \begin{equation}\label{eq:29}
        F(\rmi t \tau)-F(-\rmi t \tau) = 8 \rmi  \kappa_m \frac{\Li{2}{r^2_{0,2}}}{\mu_0+1}t \tau + \rmi \pi \frac{\beta}{2} \sum_{k=1,2} r_{0,k} t^2 \tau^2+O(\tau^3).
    \end{equation}
    As a result, instead of~\eref{eq:15} in the lowest order in $\tau$ one finds
    \begin{equation}\label{eq:30}
        \Dt \F(a,T) = - \frac{\hbar c}{32 \pi^2 a^3} \left[\frac{\kappa_m\Li{2}{r^2_{0,2}}}{3\left(\mu_0+1\right)}\tau^2 + O\left(\tau^3\right)\right].
    \end{equation}
    It is seen that now the perturbation expansion starts from the second-order term and, thus, the thermal correction~\eref{eq:30} depends on separation distance. This gives the possibility to obtain the dominant term of the thermal correction to the Casimir pressure directly from~\eref{eq:30} using equation~\eref{eq:21}
    \begin{equation}\label{eq:31}
        \Dt P(a,T) =
        -\frac{\hbar c^2}{96 \pi^2 a^5} \left[\frac{\Li{2}{r^2_{0,2}}}{\omega_m\left(\mu_0+1\right)}\tau^2 +
        O\left(\tau^3\right)\right].
    \end{equation}
    The Casimir entropy per unit area is found from~\eref{eq:30} using equation~\eref{eq:16}.
    Returning to the dimensional temperature, one has in the dominant order
    \begin{equation}\label{eq:32}
        S(a,T) = \frac{\kb^2 \Li{2}{r^2_{0,2}}}{6 \hbar \omega_m \left(\mu_0+1\right)a^2} T +
        O\left(T^2\right).
    \end{equation}
    This goes to zero when the temperature vanishes, i.e., the Nernst heat theorem is satisfied.
    If to compare with~\eref{eq:24}, it is seen that now the leading order in $T$ is the first instead
    of the second. The reason is the account of the frequency dependence of the magnetic permeability.
    Thus, an account of the frequency dependence of $\mu$ results in the change of power in the power-type
    law like this holds for $\epsilon$\cite{r23,r25}.
\subsection{Influence of dc conductivity}
    It is common knowledge that all dielectric materials at any nonzero temperature contain some small
    concentration of free charge carriers. As a result, the dielectric permittivity $\epsilon_l$ calculated at the imaginary Matsubara frequencies should be replaced with
    \begin{equation}\label{eq:33}
        \widetilde{\epsilon}_l = \epsilon_l + \frac{4 \pi \sigma_0(T)}{\xi_l} \equiv \epsilon_l + \frac{\beta(T)}{l},
    \end{equation}
    where $\sigma_0(T)$ is the static (dc) conductivity of the dielectric and $\beta =  2\hbar \sigma_0/(\kb T)$. In previous sections we have neglected by the effect of dc conductivity in magnetodielectric materials and considered only $\epsilon_l$ with a finite value of $\epsilon_0$. It is justified by the fact that for all $l \ge 1$ it holds
    \begin{equation}\label{eq:34}
        \epsilon_l \gg \frac{4\pi \sigma_0(T)}{\xi_l}.
    \end{equation}
    For example, for quartz glass (SiO$_2$) at $T=300\mathrm{K}$ one has $\beta \sim 10^{-12}$.

    Furthermore, $\sigma_0(T) \sim \exp(-b/T)$, where $b$ is some constant different for different
    materials, and, thus, $\sigma_0(T) \to 0$ when $T \to 0$\cite{r45}. For $l=0$, however,
    $\widetilde{\epsilon}_0$ diverges. Because of this, the impact of dc conductivity on the Casimir
    free energy of magnetodielectrics is determined by the zero-frequency turn of the Lifshitz
    formula~\eref{eq:01}.

    To find this impact, we notate as $\widetilde{\F}(a,T)$ the Casimir free energy per unit
    area~\eref{eq:01} calculated using the dielectric permittivity $\widetilde{\epsilon}_l$
    and magnetic permeability $\mu_l$. From equation~\eref{eq:01} it is evident that
    \begin{equation}\label{eq:35}
    \eqalign{
        \widetilde{\F}(a,T)=\widetilde{\F}_{l=0}(a,T)+\widetilde{\F}_{l\ge1}(a,T),\\
        \F(a,T)= \F_{l=0}(a,T) +\F_{l\ge1}(a,T),
    }
    \end{equation}
    where we have separated the contributions of the zero Matsubara frequency in the free energies
    calculated using $\widetilde{\epsilon}_l$ and $\epsilon_l$. It can be shown\cite{r23} that
    \begin{equation}\label{eq:36}
        \widetilde{\F}_{l\ge1}(a,T)-\F_{l\ge1}(a,T) \equiv Q(a,T) \sim \rme^{-b/T}
    \end{equation}
    independently of the value of magnetic permeability $\mu$. Calculating the zero-frequency
    contributions to~\eref{eq:01} with account of~\eref{eq:02} and~\eref{eq:33}, we obtain
    \begin{equation}\label{eq:37}
    \eqalign{
        \F_{l=0}(a,T)=-\frac{\kb T}{16 \pi a^2} \left[\Li{3}{r^2_{0,1}}+ \Li{3}{r^2_{0,2}}\right],\\
        \widetilde{\F}_{l=0}(a,T)=-\frac{\kb T}{16 \pi a^2} \left[\zeta(3)+ \Li{3}{r^2_{0,2}}\right].
    }
    \end{equation}
    Subtracting the two equations in~\eref{eq:35} and using~\eref{eq:36} and~\eref{eq:37}, one finds
    \begin{equation}\label{eq:38}
        \widetilde{\F}(a,T) = \F(a,T) - \frac{\kb T}{16 \pi a^2} \left[\zeta(3)-\Li{3}{r^2_{0,1}}\right] + Q(a,T).
    \end{equation}
    Note that contributions of the magnetic permeability in the zero frequency terms~\eref{eq:37} have been
    canceled.

    Differentiating the Casimir free energy~\eref{eq:38} with respect to $T$ and taking the negative sign,
    we arrive to the Casimir entropy per unit area with account of the effect of dc conductivity
    \begin{equation}\label{eq:39}
        \widetilde{S}(a,T) = S(a,T) + \frac{\kb}{16 \pi a^2} \left[\zeta(3)-\Li{3}{r^2_{0,1}}\right] -
        \frac{\partial Q(a,T)}{\partial T},
    \end{equation}
    where $S(a,T)$ is defind in~\eref{eq:32}. Taking into consideration that
    \begin{equation}\label{eq:40}
        S(a,T) \to 0, \, \frac{\partial Q(a,T)}{\partial T} \to 0 \qquad \mathrm{when} \, T\to 0,
    \end{equation}
    one obtains at $T=0$
    \begin{equation}\label{eq:41}
        \widetilde{S}(a,0)=\frac{\kb}{16 \pi a^2} \left[\zeta(3)-\Li{3}{r^2_{0,1}}\right]>0.
    \end{equation}
    This means the violation of the Nernst heat theorem because the right-hand
    side of~\eref{eq:41} depends on the parameters of a system under
    consideration (the separation distance between the plates and their
    dielectric permittivity). Thus, if the  dc conductivity of magnetodielectric
    materials is taken into consideration, the Lifshitz theory violates the third
    law of thermodynamics in the same way as for ordinary dielectrics which do
    not possess magnetic properties\cite{r04,r07,r23,r24,r25}.
\section{Conclusions and discussion}\label{sec:concl}
    In this paper, we have investigated the low-temperature behavior of the Casimir
    entropy for two parallel plates made of magnetodielectric materials. For this
    purpose, the Casimir free energy at low temperature was found analytically for
    both constant and frequency-dependent dielectric permittivity and magnetic
    permeability in the form of perturbation expansions. The effect of dc
    conductivity was also taken into account. It was  shown that for magnetodielectric materials
    characterized by the finite static dielectric permittivity and magnetic permeability the Casimir
    entropy goes to zero when
    the temperature  vanishes, i.e., the third
    law of thermodynamics (the Nernst heat theorem) is satisfied. By contrast, if
    the dc conductivity is taken into account, in the limiting case of zero
    temperature the Casimir entropy goes to the positive constant
    depending on the parameters of a system, i.e., the Nernst heat theorem is violated.
    This is the same effect, as was discovered previously for
    dielectrics which
    do not possess magnetic properties\cite{r04,r07,r23,r24,r25}. In so doing, the
    value of the entropy at zero temperature does not depend on the magnetic
    permeability of the plates.

    As is mentioned in section~\ref{sec:intro}, the measurement data of several
    experiments using dielectric plates are consistent with theoretical
    predictions of the Lifshitz theory only if the contribution of dc
    conductivity is omitted in computations.  This means that experimentally
    consistent is the theoretical approach, which is also in agreement with
    thermodynamics. This would be not surprising if the dc conductivity of dielectrics
    at room temperature were not an observable physical effect contributing in many
    experiments with an exception of the Casimir effect. To conclude we can say
    that the violation of the Nernst heat theorem for magnetodielectric
    materials in the Lifshitz theory is a challenging problem which awaits for its resolution.

\ack
    The authors are grateful to V.~M.~Mostepanenko for helpful discussions.

\end{document}